\documentclass[journal,12pt,onecolumn,draftclsnofoot,letter]{IEEEtran}

\ifCLASSINFOpdf
\else

\fi

\usepackage{graphicx}
\usepackage{epstopdf}
\usepackage{balance}
\usepackage{multicol}
\usepackage{url}
\usepackage{amsmath}
\usepackage{mathtools}
\usepackage{amssymb}
\usepackage{esint}
\usepackage{amsfonts}
\usepackage{multirow}
\usepackage{wrapfig}
\usepackage{hyperref}
\usepackage{cite}
\usepackage{subcaption}
\usepackage{lipsum}
\usepackage{float}    
\usepackage{algorithm}
\usepackage{algorithmic}
\usepackage{url}
\usepackage{caption}
\usepackage[none]{hyphenat}
\usepackage[usenames,dvipsnames]{color}

\begin{document}
\title{\LARGE Federated Learning for Localization: A Privacy-Preserving Crowdsourcing Method}
\author{\IEEEauthorblockN{Bekir Sait Ciftler,
Abdullatif Albaseer, Noureddine Lasla, and Mohamed Abdallah\\}
\IEEEauthorblockA{Division of Information and Computing Technology,\\ College of Science and Engineering,\\
Hamad Bin Khalifa University, Doha, Qatar\\
\{bciftler, amalbaseer, nlasla, moabdallah\}@hbku.edu.qa}\vspace{-8mm}}

\maketitle
\begin{abstract}
Received Signal Strength (RSS) fingerprint-based localization has attracted a lot of research effort and cultivated many commercial applications of location-based services due to its low cost and ease of implementation.
Many studies are exploring the use of deep learning (DL) algorithms for localization.
DL's ability to extract features and to classify autonomously makes it an attractive solution for fingerprint based localization.
These solutions require frequent retraining of DL models with vast amounts of measurements.
Although crowdsourcing is an excellent way to gather immense amounts of data, it jeopardizes the privacy of participants, as it requires to collect labeled data at a centralized server. 
Recently, federated learning has emerged as a practical concept in solving the privacy preservation issue of crowdsourcing participants by performing model training at the edge devices in a decentralized manner; the participants do not expose their data anymore to a centralized server.
This paper presents a novel method utilizing federated learning to improve the accuracy of RSS fingerprint-based localization while preserving the privacy of the crowdsourcing participants.
Employing federated learning allows ensuring \emph{preserving the privacy of user data} while enabling an adequate localization performance with experimental data captured in real-world settings.
The proposed method improved localization accuracy by $1.8$ meters when used as a booster for centralized learning and achieved satisfactory localization accuracy when used standalone.
\end{abstract}
\begin{IEEEkeywords}
Crowdsourcing, Deep Learning, Federated Learning, Fingerprint, Indoor Localization, Neural Networks, Privacy-Preserving, Received Signal Strength.
\end{IEEEkeywords}
\section{Introduction}
\label{sect:Introduction}
Advances in computational power and parallel processing abilities, as well as the availability of the massive amount of data, created considerable interest in deep learning (DL) in many fields of wireless communications research\cite{Zhang2019DeepLearning}.
The proliferation of smart devices also enabled a wide range of location-based services that relies on DL-based Indoor Positioning Systems (IPS) utilizing sensory data gathered from the smart devices\cite{Zafari2019Survey}.
These systems require an immense amount of data for training.
The data gathering process is time-consuming and strenuous.
Data crowdsourcing has recently gained attention to alleviate the difficulty of data gathering by allowing public users to contribute to datasets in a participatory sensing manner.

Crowdsourcing is an essential solution for creating large datasets for DL for IPS, which is also used in various applications such as image recognition, medical research, and market research.
It is used to reduce the required effort for implementing DL-based IPS\cite{Wang2016IndoorS}.
The traditional crowdsourcing technique relies on the users who send their data directly to the centralized server, which violates users' privacy.
There are methods for handling privacy-related problems such as data anonymization, encrypted evaluation, and noise injection. However, methods such as encrypted evaluation and anonymization are challenging to implement in real-world applications due to computational complexity.
Additionally, methods similar to noise injection may cause long term degradation in the performance of DL-based applications.
Recently, Federated Learning is an emerging method that enables privacy-preserving deep learning while keeping the benefits of crowdsourcing\cite{konen2016federated}.

Federated learning has gathered a tremendous interest recently due to its privacy-preserving nature as well as the efficient use of resources by utilizing the processing power of edge devices\cite{niknam2019federated}.
The local models are trained with user data at the device, and only models which are smaller in size are shared with the centralized server\cite{mcmahan2016communicationefficient}.
Hence, federated learning is also communication-wise efficient since it avoids the transmission of vast amounts of data\cite{park2018wireless}.

\begin{figure}[t]
    \centering
    \includegraphics[width=0.45\textwidth]{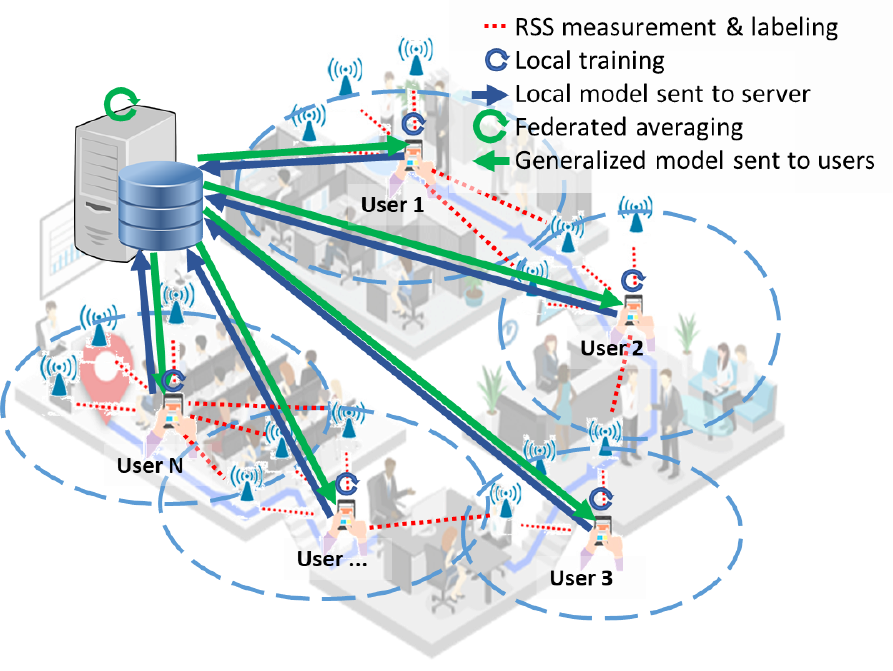}
    \caption{Federated learning-based crowdsourcing system.}
    \label{fig:systemmodel}
    \vspace{-5mm}
\end{figure}

In this paper, we propose a novel method utilizing federated learning to create an RSS fingerprint-based localization crowdsourcing system, as illustrated in Fig.~\ref{fig:systemmodel}.
Our main aim is creating a localization system employing crowdsourcing without violating user privacy.
We designed a multilayer perceptron (MLP) as a global training model to extract location information from RSS features.
We utilized TensorFlow and Keras libraries for simulating our federated learning approach using the UJIndoorLoc database\cite{Ujiloc2014Torres}.
Numerical results show that our proposed method is a viable solution for crowdsourcing for RSS fingerprint-based localization systems.

Compared to existing DL-based indoor localization systems, we list our main contributions as below:
\begin{itemize}
    \item The main contribution of this work is enabling \emph{a privacy-preserving crowdsourcing technique} for RSS fingerprint-based localization with keeping local data where it is generated and conveying only local models throughout the learning process.
    \item Secondly, various \emph{heterogeneity models are used for unbalanced local data sizes and non-IID spatial distribution} of user data in three scenarios.
    These scenarios show the effect of unbalanced and non-IID data on the localization accuracy and the convergence time of federated learning.
    \item Thirdly, we test our proposed method using experimental data captured in real-world settings\cite{Ujiloc2014Torres}, which has localization Mean Absolute Error (MAE) up to below $5$ meters in $390$ m x $270$ m area.
    \item Finally, we demonstrate that federated learning can perform localization accuracy (up to below $1$ meters) close to centralized training without violating user privacy, and it can \emph{boost performance of centralized training} reducing MAE by $1.8$ meters.
\end{itemize}

This paper is structured as follows.
In Section~\ref{sect:RelatedWork}, a brief overview of existing literature on traditional machine learning-based fingerprint localization techniques are presented. 
We explain our proposed federated learning for RSS fingerprint-based localization technique in Section~\ref{sect:SystemModel}.
Subsequently, we analyze the numerical results for three comprehensive scenarios in Section~\ref{sect:Results}.
Finally, we present our concluding remarks in Section~\ref{sect:Conclusion}.

\section{Literature Review}
\label{sect:RelatedWork}
Fingerprinting-based localization is highly popular due to ease of implementation and high accuracy.
There are several traditional fingerprinting applications, such as \cite{youssef2005horus}, where the system employs a probabilistic approach utilizing the RSS histogram of the access points (APs) for location clustering.
Another well-known technique is a statistical approach based on k Nearest Neighbors (kNN).
In this approach, the differences in the RSS values with various distance metrics for corresponding reference points (RP) in the database.
The RP with the smallest difference for the distance metric is estimated as the location of the user\cite{Xie2016KNN}.

Localization research utilizing deep learning has gained momentum with the availability of vast amounts of data incorporating location-related information gathered from smart devices as well as an increase in computational power.
The authors propose a method called DeepFi based on DL using channel state information (CSI) fingerprint as input in~\cite{Wang2017CSIDeep}, and they compared their method and shown excellence to traditional methods such as HORUS\cite{youssef2005horus} and maximum likelihood\cite{brunato2005statistical}.
The proposed DeepFi method outperformed several traditional RSS and CSI based schemes in two representative indoor environments.
The authors subsequently increased the accuracy utilizing calibrated phase information extracted from CSI\cite{Wang2016Phase} as the fingerprint.
In another study, a Convolutional Neural Network (CNN) model is employed utilizing the angle of arrival information\cite{Wang2018DCNN}.
However, each of these techniques requires hardware modifications on user devices and gathering sensitive user data for centralized training.

All of the traditional localization techniques are based on centralized training models, and they require the transmission of sensitive data from users, jeopardizing user privacy.
However, the increasing awareness about privacy and the policies for rules and regulations from legislative entities are restricting the reach of those methods to the sensitive data.
Hence a solution, such as federated learning, is not only viable but also required for future localization systems.

Federated learning is an emerging solution for privacy-preserving distributed learning\cite{konen2016federated} since it is an answer to several shortcomings of traditional machine learning algorithms.
Federated learning allows multiple users to train their models locally without revealing their sensitive data to the centralized unit.
It is communication-wise efficient since models are significantly smaller than the raw data in most of the applications\cite{mcmahan2016communicationefficient}.
Federated learning is proven to be robust against unbalanced and non-IID data.
It is applicable to a variety of deep learning methods such as MLP, CNN, and Long-Short-Term-Memory Recurrent Neural Networks.
In our proposed method, we use MLP architecture with federated learning for the RSS Fingperint-based localization system to overcome problems of traditional deep learning-based localization systems.
\section{Privacy-Preserving Localization with Federated Learning}
\label{sect:SystemModel}

\begin{figure}[t]
    \centering
    \includegraphics[width=0.3\textwidth]{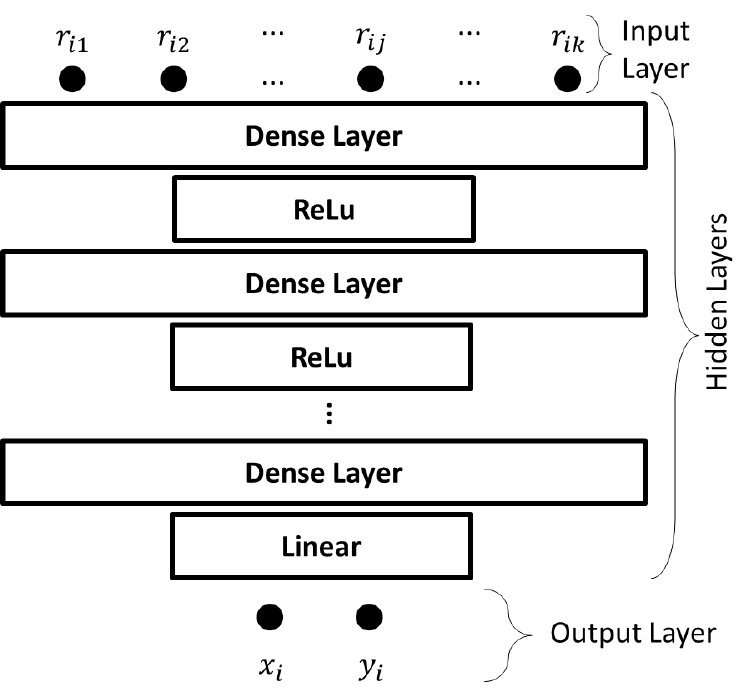}
    \caption{MultiLayer Perceptron global model.}
    \label{fig:MLP}
    \vspace{-5mm}
\end{figure}

\subsection{Data Preprocessing}
\label{subsect:DataProc}
In this paper, we consider a privacy-preserving federated learning-based \emph{crowdsourcing} system for RSS Fingerprint-based localization for an indoor scenario, as seen in Fig.\ref{fig:systemmodel}.
The initial global MLP model is generated based on the training data gathered for the centralized server to determine the number of inputs $k$, which is the number of fixed APs in the building, as in Fig.~\ref{fig:MLP}.
Initially, a set of $m$ training samples of RSS measurements are collected from a set of $k$ fixed APs, broadcasting WiFi beacons in the building, and a set of $n$ users gathering RSS measurements from received beacons.
The labeling of $i$th sample with the location information is done in the cartesian coordinates of the RSS measurement and represented as 
\begin{equation}
    \boldsymbol{y_i}=[x_i~y_i]^T,~\forall i \in M.\label{eq:labelvector}
\end{equation}

Each training sample $i$ includes $l_i$ RSS values from a subset APs of $k$ APs (i.e., $l_i \leq k$) since not all APs are in the coverage of the user at the same time.
The RSS value of beacon from $j$th AP for $i$th training sample is represented with $r_{ij}$. 
Hence the measurement vector for $i$th sample can be defined as
\begin{align}
\boldsymbol{s_i} = [r_{ij_1}~\cdots~r_{ij}~\cdots~r_{ij_{l_i}}],\forall i \in M,~j\in L_i,\label{eq:measurementvector}
\end{align}
where $M$ is the set of all training samples, and $L_i$ is the subset of APs with an RSS value in the $i$th training sample. 

Thus, the initial MLP model has $k$ nodes in the input layer, which involves the vector of RSS values from WiFi beacons.
In case the beacon from any AP is not available (that AP is not in the coverage), the RSS value of the corresponding AP is set to a constant value ($C$) as a missing value.
The constant for missing value is set to a lower value than the minimum RSS possible, which represents it is not in the coverage.
In our simulations, we used $C=-150$ to represent missing RSS values.
Hence the below equation represents the input vector for MLP model for a sample $i$
\begin{align}
    \boldsymbol{x_i} = [r_{i1}~\cdots~r_{ij}~\cdots~r_{ik}],\forall i \in M,\label{eq:inputvector}
\end{align}
where
\begin{align}
    r_{ij}=C,~\forall j~\notin~L_i.
\end{align}

\subsection{Global MLP Model and Centralized Training}
The global model has a defined amount number of hidden nodes in specified numbers of hidden (dense) layers.
After the generation of the global model, it is trained with existing $m$ samples.
The MLP training can be shown as a minimization problem considering loss function $f$ as below
\begin{align}
    \min \limits_{\boldsymbol{w}}\frac{1}{m}\sum_{i=1}^{m} f(\boldsymbol{w},\boldsymbol{x_i},\boldsymbol{y_i}),\label{eq:minimization}
\end{align}
where $\boldsymbol{w}$ is the model weights, $m$ is the number of training samples, $\boldsymbol{x_i}$ is the $i$th input training sample, and $\boldsymbol{y_i}$ is the label for $i$th sample.
The global MLP model is trained in the centralized server.
The following step is transmitting the global MLP model to the users for federated learning, as described in the next subsection.

\subsection{Federated Learning}
\label{subsect:FedLearn}

\begin{figure}[t]
    \centering
    \includegraphics[width=0.2\textwidth]{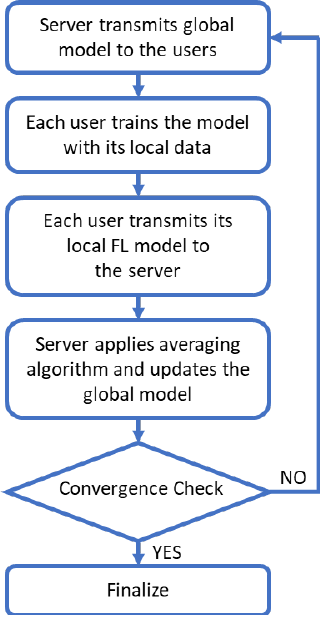}
    \caption{Federated Learning procedure.}
    \label{fig:algorithm}
    \vspace{-5mm}
\end{figure}

The general procedure for federated learning is summarized in Fig.~\ref{fig:algorithm}.
The general model is transmitted to the users by the centralized server.
In each round, $n$ users gather training data (RSS measurements) from APs in the vicinity and label them with corresponding measurement locations in different areas of the building.
These users may change their positions; hence each measurement may consist beacons from different APs nearby at each sample.
Subsequently, the users train the general model with their data, considering the minimization problem given in \eqref{eq:minimization}.
Each user conveys its local model and its number of training samples to the centralized server.
The centralized server applies the federated weighted averaging and updates the global model based on local models of users as below
\begin{align}
    \boldsymbol{w^{t+1}}&=\frac{1}{H^t}\sum_{u=1}^{N}m^t_u\boldsymbol{w^t_u},\label{eq:weightedaveraging}\\
    H^t&=\sum_{u=1}^{n}m^t_u,
\end{align}
where $H$ is the total number of training samples, $t$ is the round.
After federated averaging, the global model is updated and transmitted back to the edge devices.
The edge devices retrain the global model with their local data.
These steps are repeated until convergence.

\begin{table}[t]
    \centering
    \caption{Federated Learning system parameters.}
    \begin{tabular}{cc}
        \hline
        \multicolumn{1}{|l|}{\textbf{Parameter}} & \multicolumn{1}{l|}{\textbf{Value}} \\ \hline\hline
        \multicolumn{1}{|l|}{Deep Learning Libraries} & \multicolumn{1}{l|}{Google TensorFlow \& Keras} \\ \hline
        \multicolumn{1}{|l|}{Optimizer} & \multicolumn{1}{l|}{Adam\cite{kingma2014adam}} \\ \hline
        \multicolumn{1}{|l|}{Learning Rate} & \multicolumn{1}{l|}{$0.0001$} \\ \hline
        \multicolumn{1}{|l|}{$\beta_1$, $\beta_2$} & \multicolumn{1}{l|}{$0.1$, $0.99$} \\ \hline
        \multicolumn{1}{|l|}{Hidden Layer Formation} & \multicolumn{1}{l|}{$20$ x $10$ x $10$ x $10$} \\ \hline
        \multicolumn{1}{|l|}{Activation Function} & \multicolumn{1}{l|}{ReLu} \\ \hline
        \multicolumn{1}{|l|}{Loss Function} & \multicolumn{1}{l|}{Mean Absolute Error} \\ \hline
        \multicolumn{1}{|l|}{{Batch Size}} & \multicolumn{1}{l|}{{$100$}}  \\ \hline
        \multicolumn{1}{|l|}{{Number of Epochs in a Round}} & \multicolumn{1}{l|}{{$10$}}  \\ \hline
        \multicolumn{1}{|l|}{{Number of Access Points (k)}} & \multicolumn{1}{l|}{{$520$}}  \\ \hline
        \multicolumn{1}{|l|}{{Number of Users (n)}} & \multicolumn{1}{l|}{{$1$ to $15$}}  \\ \hline
        \multicolumn{1}{|l|}{{Number of Training Samples}} & \multicolumn{1}{l|}{{$3000$ to $15000$}}  \\ \hline
        \multicolumn{1}{|l|}{{Number of Test Samples}} & \multicolumn{1}{l|}{{$3000$}}  \\ \hline
        \label{table:parameter}
    \end{tabular}
    \vspace{-5mm}
\end{table}

\section{Numerical Results}
\label{sect:Results}

In this section, the performance of the proposed crowdsourcing method is benchmarked to a centralized DL method in various settings and scenarios.
Publicly available RSS data from UJIIndoorLoc database\cite{Ujiloc2014Torres} is used to evaluate our proposed federated learning-based localization system.
The UJIIndoorLoc database includes $19937$ samples distributed over $390$ meters in length to $270$ meters in width area over four floors for training and testing.
It has RSS values from 520 distinct APs gathered labeled with timestamp and location information in longitude and latitude.

The simulation parameters for federated learning are summarized in Table~\ref{table:parameter}.
TensorFlow and Keras libraries are utilized to implement the proposed method.
The optimizer is selected as Adam\cite{kingma2014adam}.
The learning rate and optimizer parameters are found with heuristic search and are as shown in Table~\ref{table:parameter}.
The MLP layers are created as $20x10x10x10$ due to the processing power and energy limitations of edge devices.
More hidden layers with a larger number of hidden nodes would allow better training and feature extraction of RSS measurements. However, it would be power consuming and computationally overloading for the edge devices.
The use of federated learning with this MLP formation would not bear a significant computational or energy burden on today's powerful, Neural Processing Unit-integrated chipsets such as Snapdragon 845 or Kirin 970\cite{ignatov2018ai}.
Batch size and the number of epochs for MLP learning are set to $100$ and $10$ per round.
Three generic scenarios are simulated with UJIIndoorLoc database\cite{Ujiloc2014Torres} to prove the merits of our methodology.
The test dataset has $4000$ samples in all scenarios uniformly distributed over the building, assumed to represent a real use case.

\subsubsection{Scenario-1} 
As discussed in \ref{subsect:FedLearn}, the user data is not transmitted in federated learning, instead the local models are trained.
In this scenario, we show the prominence of using federated learning for crowdsourcing of RSS fingerprint for localization.
The centralized server has $3000$ training samples distributed uniformly over the building in its dataset.
The centralized training utilizes the centralized database due to the privacy concerns of users.
In the federated learning, each user provides its local model trained by their $500$ training samples in each round to the centralized server.
The centralized server applies federated averaging as in \eqref{eq:weightedaveraging}, including its centralized model.
The localization accuracy increases with additional users due to the utilization of local datasets through federated learning.

\subsubsection{Scenario-2} The performance of the proposed method is benchmarked against the centralized training method in this scenario.
We divide the uniformly distributed dataset into $1000$ training samples per user, and the federated learning is applied as discussed in Section~\ref{subsect:FedLearn}.
Separately, the centralized server has the same user data with federated learning and combined them at its database for centralized training.
Hence, the performance of federated learning and centralized learning is compared with the same dataset.

\subsubsection{Scenario-3} The functionality of the proposed method is tested against both location and training size heterogeneity, similar to real-world applications.
In this realistic scenario, instead of the same amount of data per user as tested in Scenario-2, the amount of data per user is a random variable with a Gaussian distribution with $1000$ samples as a mean and $500$ as standard deviation.
Additionally, these samples are spatially divided; thus, each user covers a different exclusive region of the building.
Each user trains its local model for a different region, which makes the convergence time of the global model longer.
Hence, there are two realistic challenges against federated learning; user data size heterogeneity and sample location heterogeneity.

\begin{table}[t]
    \centering
    \caption{MAE (m) at round t=100 for various scenarios.}
    \begin{tabular}{ccccc}
        \hline
        \multicolumn{1}{|l|}{\textbf{Scenario}} & \multicolumn{1}{c|}{\textbf{Federated}} &  \multicolumn{1}{c|}{\textbf{Centralized}} & \multicolumn{1}{c|}{\textbf{Federated}} &
        \multicolumn{1}{c|}{\textbf{Centralized}}\\
        \multicolumn{1}{|c|}{\textbf{}} & 
        \multicolumn{1}{c|}{\textbf{Test}} &
        \multicolumn{1}{c|}{\textbf{Test}} &
        \multicolumn{1}{c|}{\textbf{Training}} &
        \multicolumn{1}{c|}{\textbf{Training}}\\\hline
        \multicolumn{1}{|l|}{\textbf{S-1 n=5}} & 
        \multicolumn{1}{c|}{{5.88}} &
        \multicolumn{1}{c|}{{6.78}} &
        \multicolumn{1}{c|}{{3.09}} &
        \multicolumn{1}{c|}{{2.43}}\\\hline
        \multicolumn{1}{|l|}{\textbf{S-1 n=10}} & 
        \multicolumn{1}{c|}{{5.55}} &
        \multicolumn{1}{c|}{{6.78}} &
        \multicolumn{1}{c|}{{2.85}} &
        \multicolumn{1}{c|}{{2.43}}\\\hline
        \multicolumn{1}{|l|}{\textbf{S-1 n=15}} & 
        \multicolumn{1}{c|}{{4.98}} &
        \multicolumn{1}{c|}{{6.78}} &
        \multicolumn{1}{c|}{{2.33}} &
        \multicolumn{1}{c|}{{2.43}}\\\hline
        \multicolumn{1}{|l|}{\textbf{S-2 n=5}} & 
        \multicolumn{1}{c|}{{8.08}} &
        \multicolumn{1}{c|}{{7.35}} &
        \multicolumn{1}{c|}{{4.52}} &
        \multicolumn{1}{c|}{{3.52}}\\\hline
        \multicolumn{1}{|l|}{\textbf{S-2 n=10}} & 
        \multicolumn{1}{c|}{{5.98}} &
        \multicolumn{1}{c|}{{4.76}} &
        \multicolumn{1}{c|}{{4.41}} &
        \multicolumn{1}{c|}{{3.41}}\\\hline
        \multicolumn{1}{|l|}{\textbf{S-2 n=15}} & 
        \multicolumn{1}{c|}{{5.25}} &
        \multicolumn{1}{c|}{{4.91}} &
        \multicolumn{1}{c|}{{4.34}} &
        \multicolumn{1}{c|}{{3.09}}\\\hline
        \multicolumn{1}{|l|}{\textbf{S-3 n=5}} & 
        \multicolumn{1}{c|}{{7.35}} &  
        \multicolumn{1}{c|}{{6.08}} &
        \multicolumn{1}{c|}{{4.46}} &
        \multicolumn{1}{c|}{{3.72}}\\\hline
        \multicolumn{1}{|l|}{\textbf{S-3 n=10}} & 
        \multicolumn{1}{c|}{{6.42}} &
        \multicolumn{1}{c|}{{4.73}} &
        \multicolumn{1}{c|}{{4.31}} &
        \multicolumn{1}{c|}{{3.28}}\\\hline
        \multicolumn{1}{|l|}{\textbf{S-3 n=15}} & 
        \multicolumn{1}{c|}{{5.53}} &
        \multicolumn{1}{c|}{{4.27}} &
        \multicolumn{1}{c|}{{3.90}} &
        \multicolumn{1}{c|}{{3.06}}\\\hline
        \label{table:results}
    \end{tabular}
    \vspace{-5mm}
\end{table}

A summary of simulation results are provided in Table~\ref{table:results}.
We used the Mean absolute error (MAE) to benchmark the performance of our proposed method. 
MAE is an essential metric for localization systems\cite{Zafari2019Survey}.
Federated learning could achieve MAE below $5$ meters in simulations, which is very reasonable considering the dimensions of the building ($390$ meters x $270$ meters) that is used for creating the dataset\cite{Ujiloc2014Torres}.
The necessity of using federated learning is presented in Scenario-1, where federated learning is used as a booster to the centralized learning.
It has shown that the accuracy is improved significantly with the contributions of local models while keeping user privacy.
The difference between centralized learning and federated learning is always lower than $1$ meter for Scenario-2, and just above $1$ meter for Scenario-3 due to its heterogeneous nature.
Federated learning achieves better results with the increasing number of users, and its performance gets closer to centralized learning.

\begin{figure}[t]
    \centering
    \begin{subfigure}[t]{0.4\textwidth}
        \centering
        \includegraphics[width=1\linewidth]{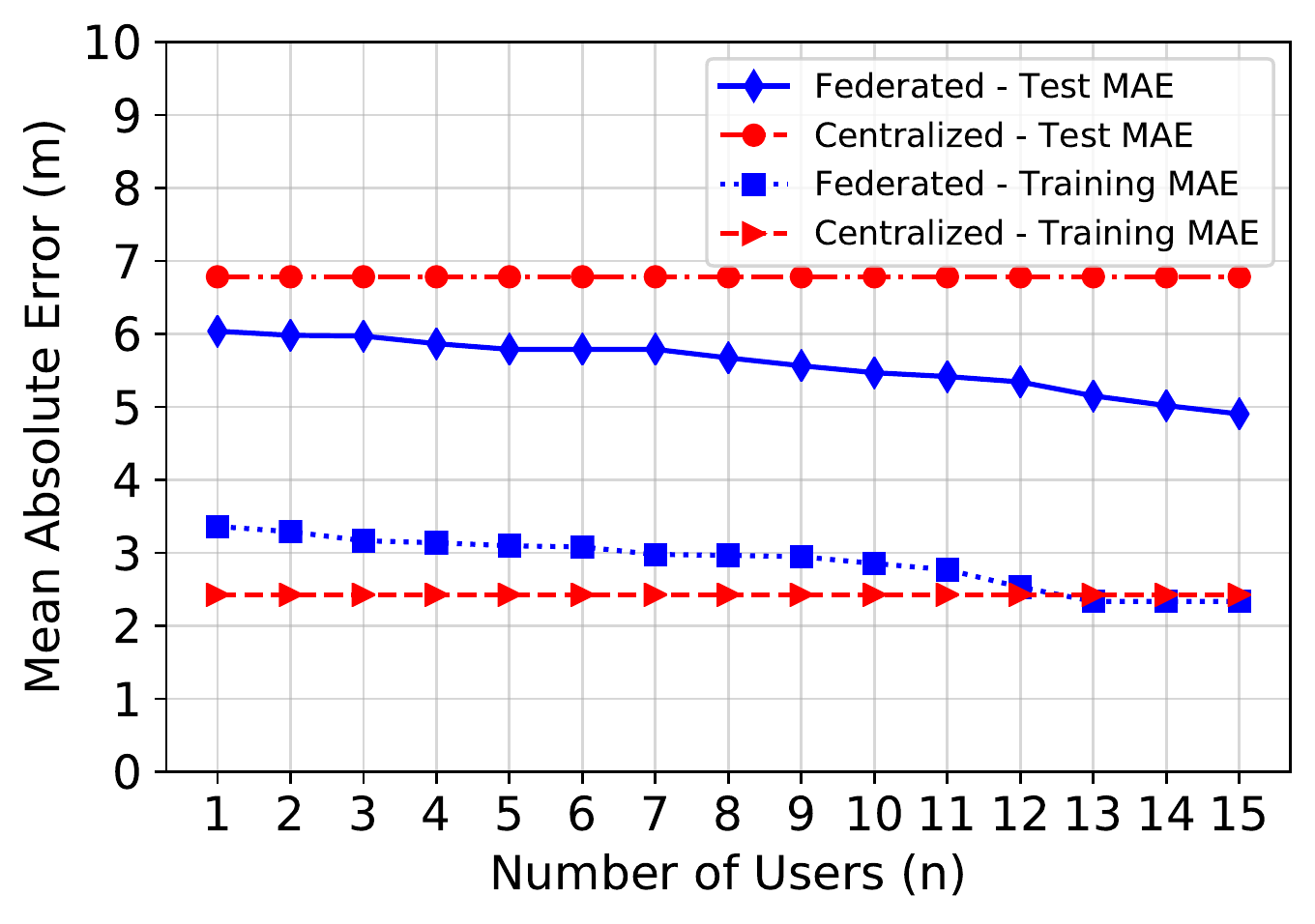}
        \caption{Scenario-1 for t=100}
    \end{subfigure}\\
    \begin{subfigure}[t]{0.4\textwidth}
        \centering
        \includegraphics[width=1\linewidth]{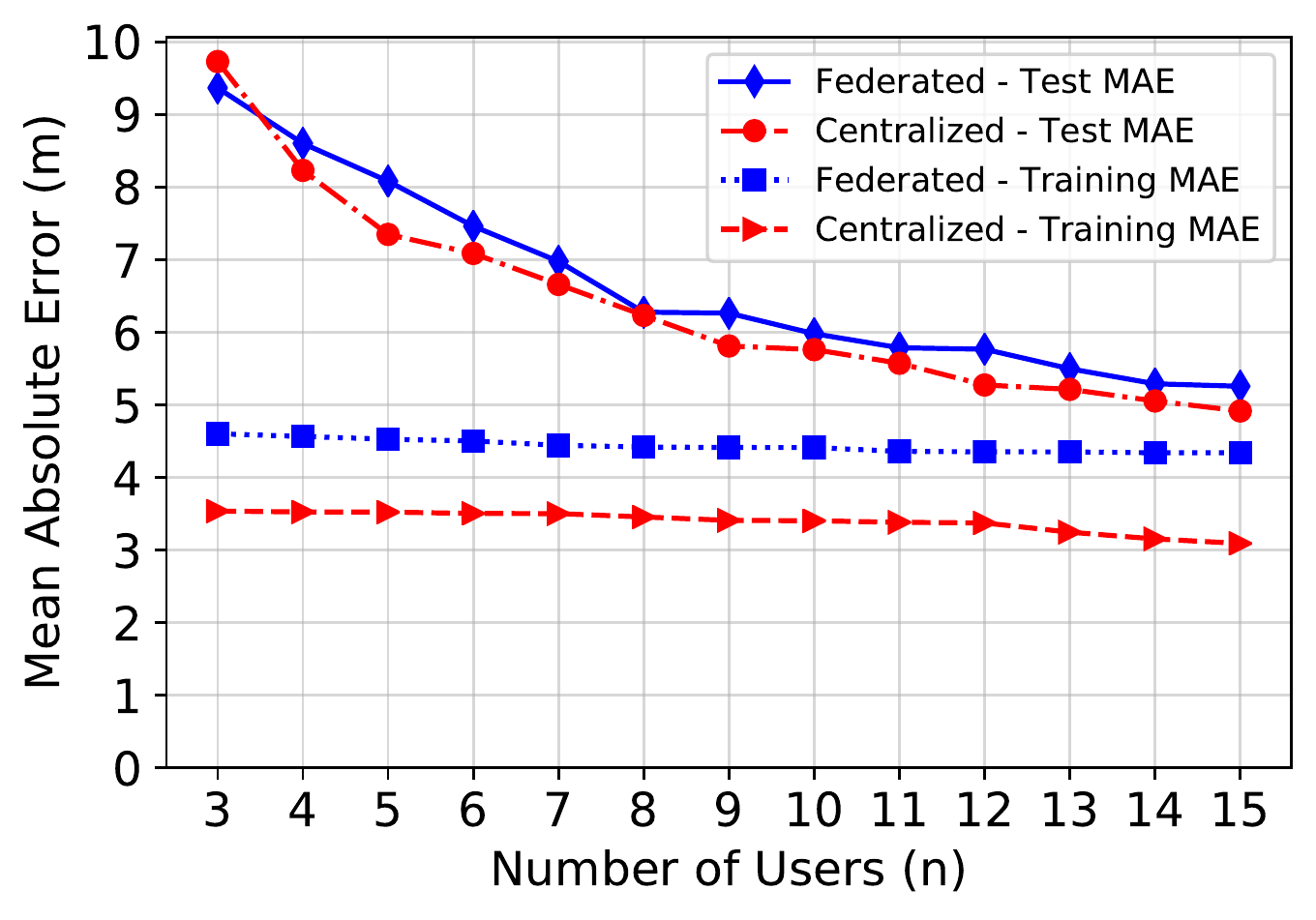}
        \caption{Scenario-2 for t=100}
    \end{subfigure}\\
    \begin{subfigure}[t]{0.4\textwidth}
        \centering
        \includegraphics[width=1\linewidth]{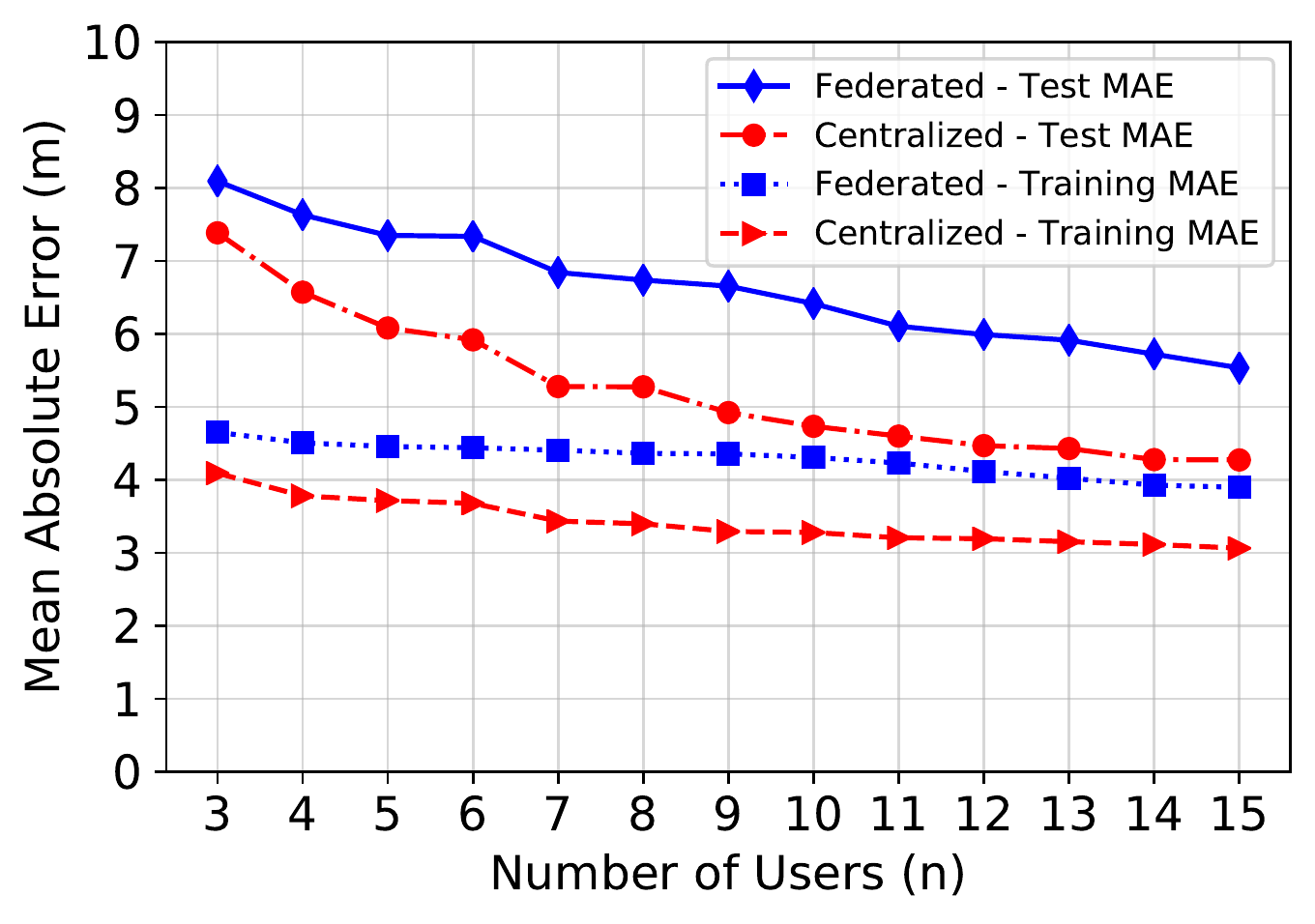}
        \caption{Scenario-3 for t=100}
    \end{subfigure}
    \caption{Effect of number of users and various heterogeneity scenarios on accuracy.}
    \label{fig:ResultsT}
    \vspace{-5mm}
\end{figure}

The effect of the number of users on federated learning localization accuracy is shown in Fig.~\ref{fig:ResultsT}.
As explained in Scenario-1, using federated learning instead of relying upon only a centralized database increases localization accuracy remarkably.
Even using a single additional contributor in federated setting drops the localization MAE by $0.65$ meters, as shown in Fig.~\ref{fig:ResultsT}(a).
Additional crowdsourcing participants notably decrease the test localization MAE.
The test localization MAE decreases by $0.9$, $1.23$, and $1.8$ meters for $n=5$, $10$, and $15$, respectively.
It is shown that the training error of the centralized learning is ($2.43$ meters) is relatively lower compared to the training error of federated method ($3.09$ meters) for five users in the system ($n=5$) for Scenario-1.
The difference between training and test results is highest in this scenario due to the smaller number of samples.
Note that the training and the test accuracy of the centralized method are constant due to using the same dataset since, in this scenario, users reject to provide their local data due to privacy concerns.

We provide a more realistic benchmark for the performance of federated learning in Fig.~\ref{fig:ResultsT}(b), considering Scenario-2.
In this scenario, we compare the performance of federated learning and centralized learning with the same sets of data.
The users waive their privacy on the data in case of centralized learning and transmit their collected data to the centralized server.
The training is done with all the user data.
In the case of federated learning, the users keep their privacy by training the model locally and keeping the data where it is generated.
The performance of the localization accuracy difference is minimal, considering the gain in the privacy of users.
The difference between centralized learning and federated learning is always below $1$ meter.

Scenario-3 has the most challenging task for the proposed federated learning-based method, as shown in Fig.\ref{fig:ResultsT}(c).
The MAE is $5.53$ meters for the federated model compared to $4.27$ meters of the centralized model for the test dataset.
One should note that the reason behind this gap is increased convergence time, and the federated model reduces the gap of less than $1$ meter after several rounds past $100$ rounds.

\begin{figure*}[t]
    \centering
    \begin{subfigure}[t]{0.32\textwidth}
        \centering
        \includegraphics[width=.95\linewidth]{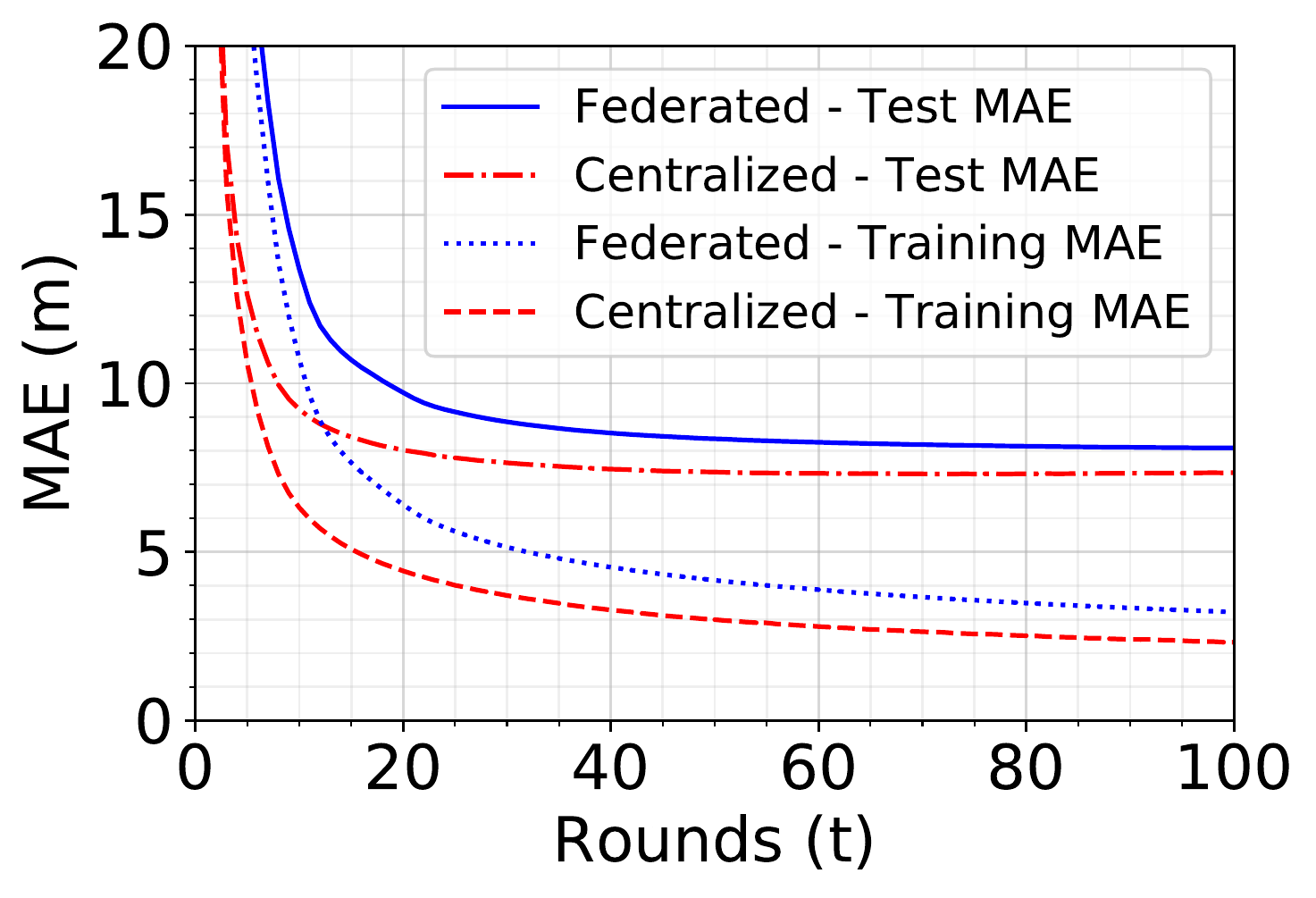}
        \caption{S-2 for n=5}
    \end{subfigure}
    \begin{subfigure}[t]{0.32\textwidth}
        \centering
        \includegraphics[width=.95\linewidth]{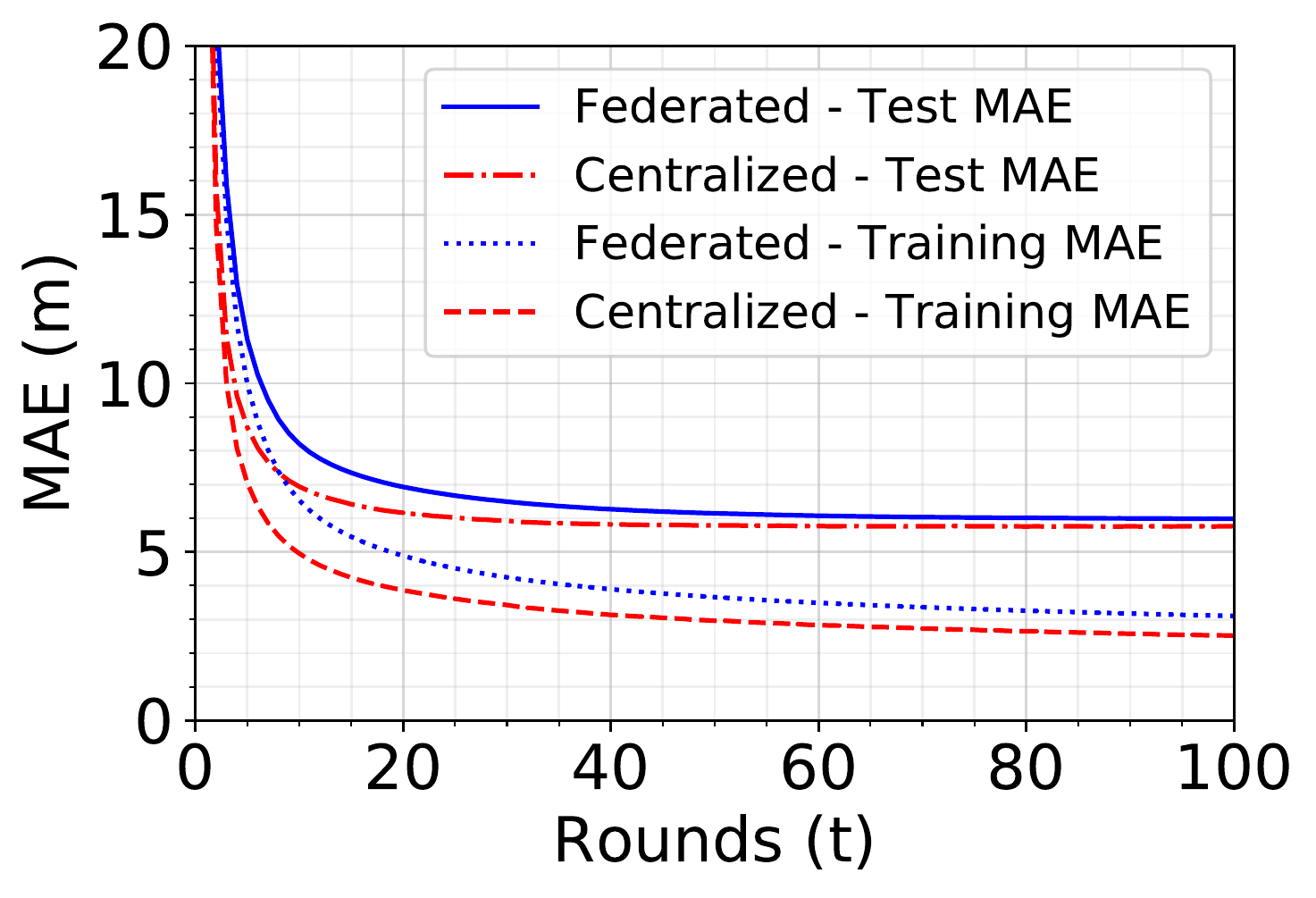}
        \caption{S-2 for n=10}    
    \end{subfigure}
    \begin{subfigure}[t]{0.32\textwidth}
        \centering
        \includegraphics[width=.95\linewidth]{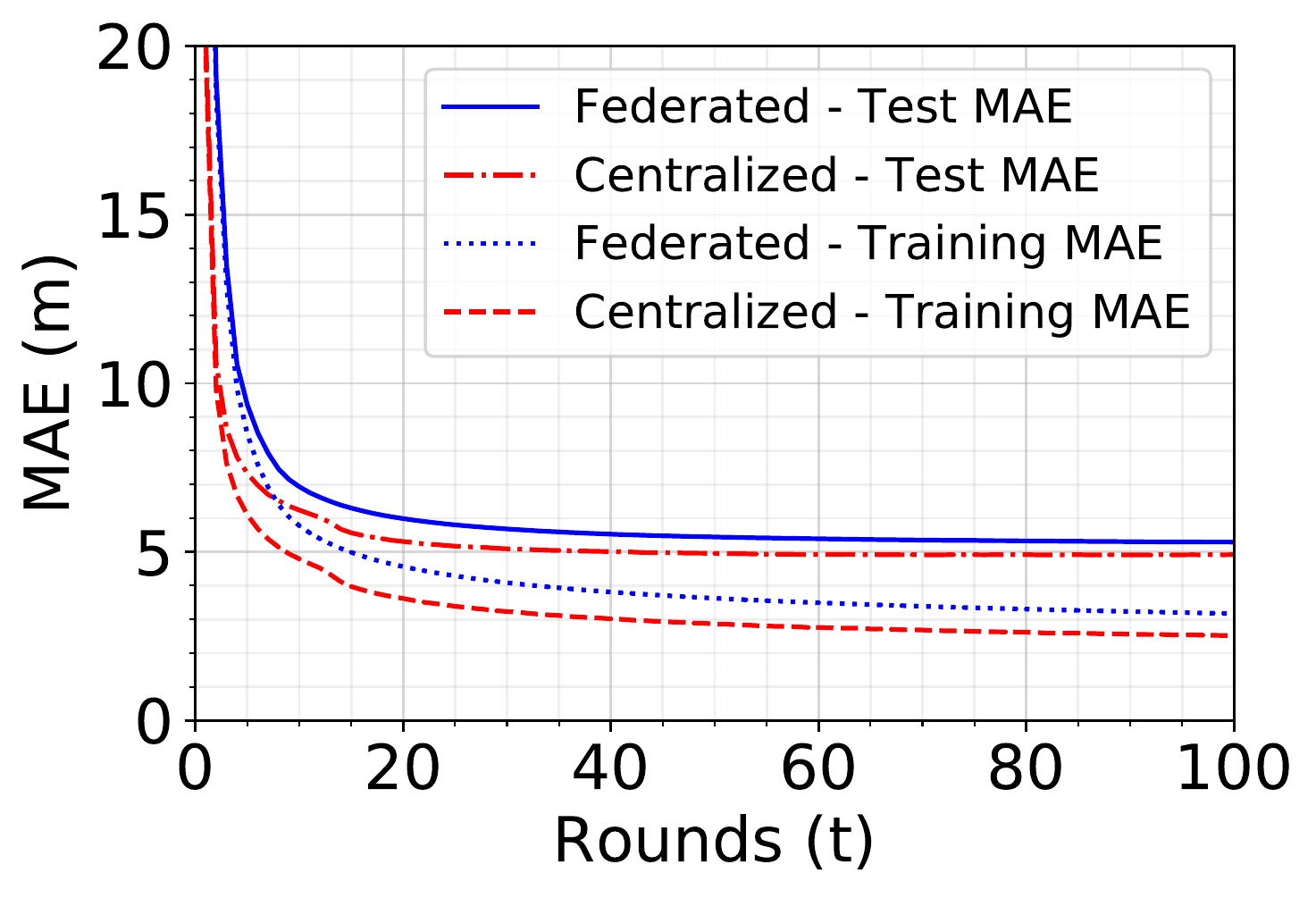}
        \caption{S-2 for n=15}
    \end{subfigure}\\
        \begin{subfigure}[t]{0.32\textwidth}
        \centering
        \includegraphics[width=.95\linewidth]{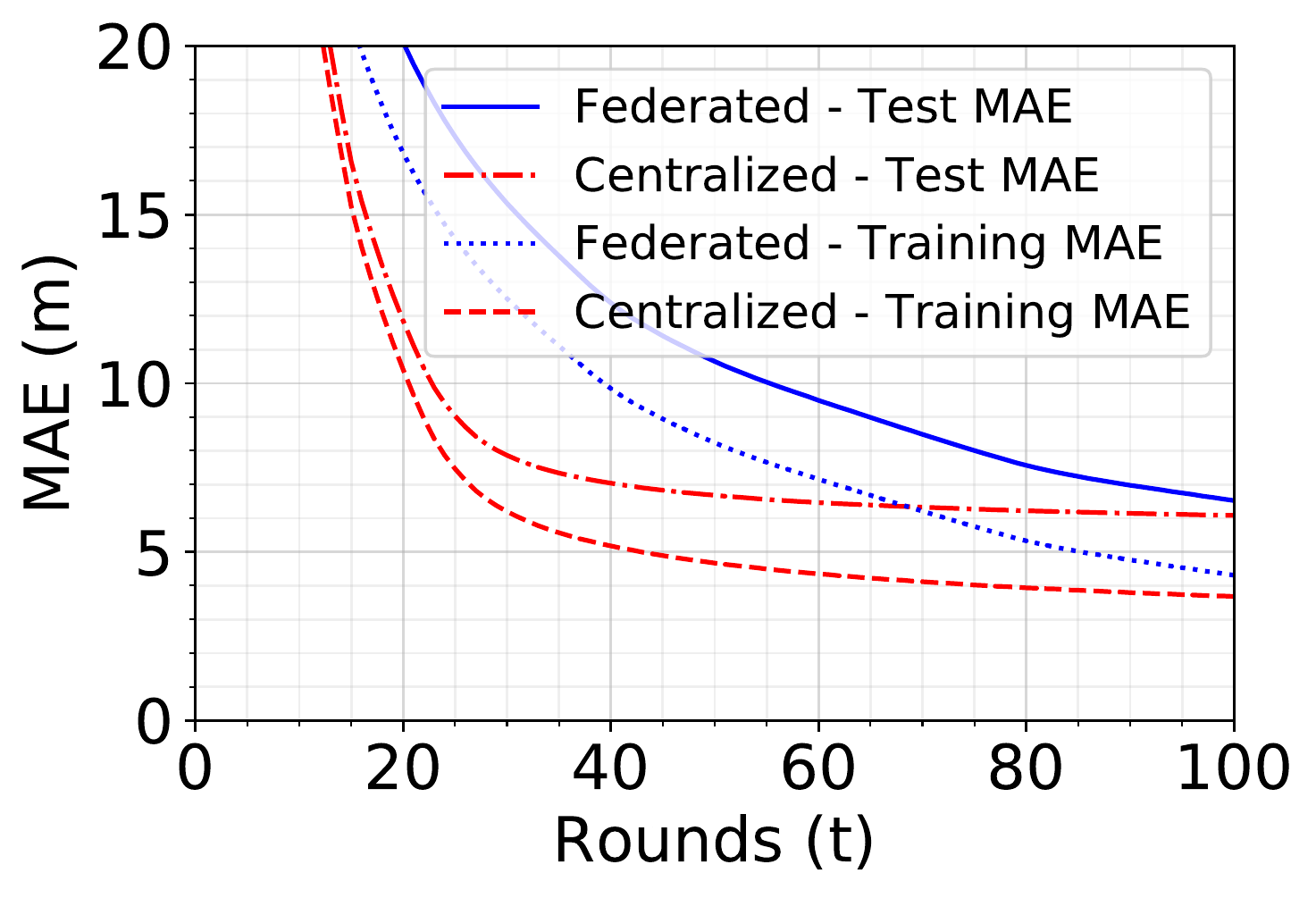}
        \caption{S-3 for n=5}
    \end{subfigure}
    \begin{subfigure}[t]{0.32\textwidth}
        \centering
        \includegraphics[width=.95\linewidth]{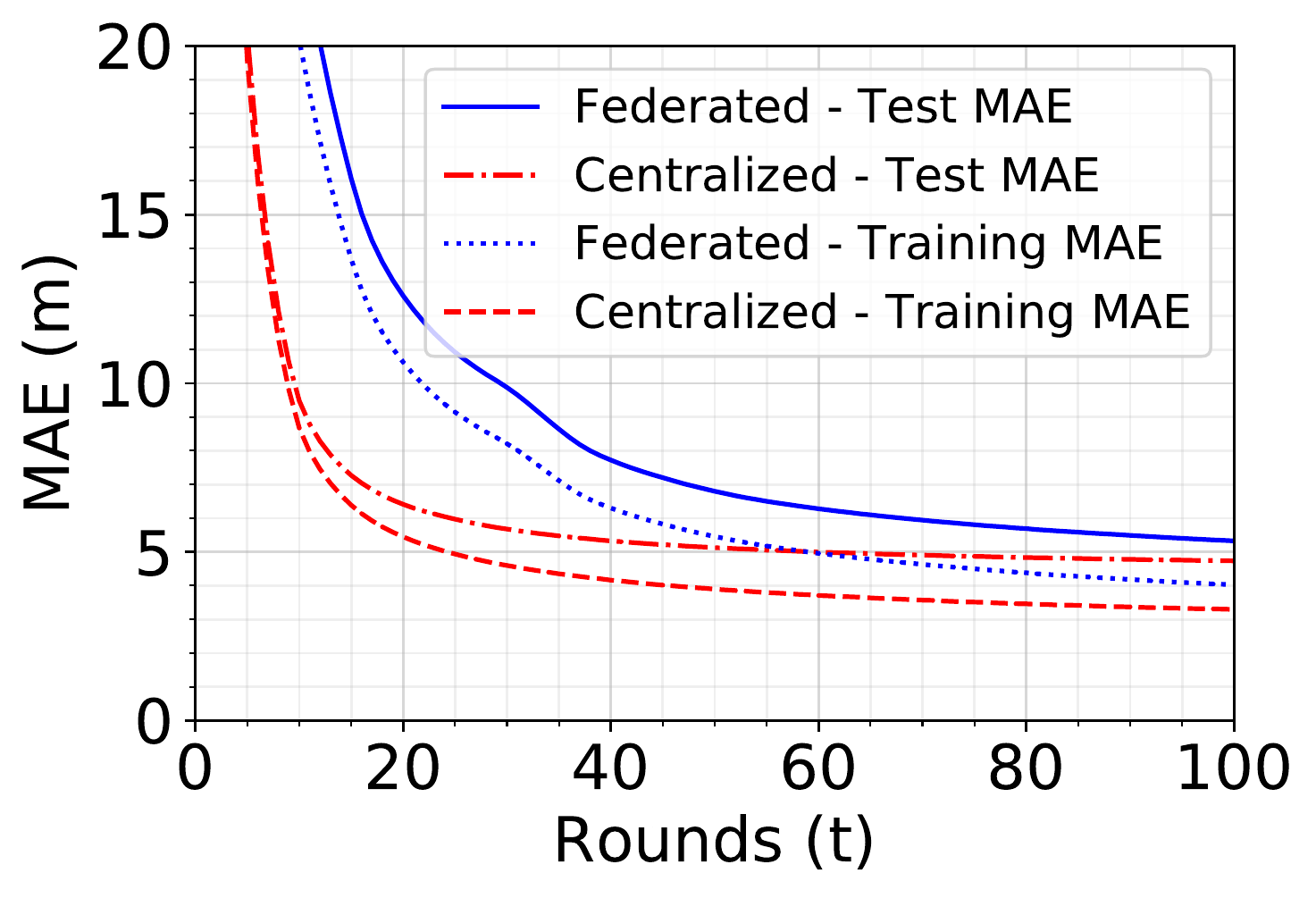}
        \caption{S-3 for n=10}
    \end{subfigure}
    \begin{subfigure}[t]{0.32\textwidth}
        \centering
        \includegraphics[width=.95\linewidth]{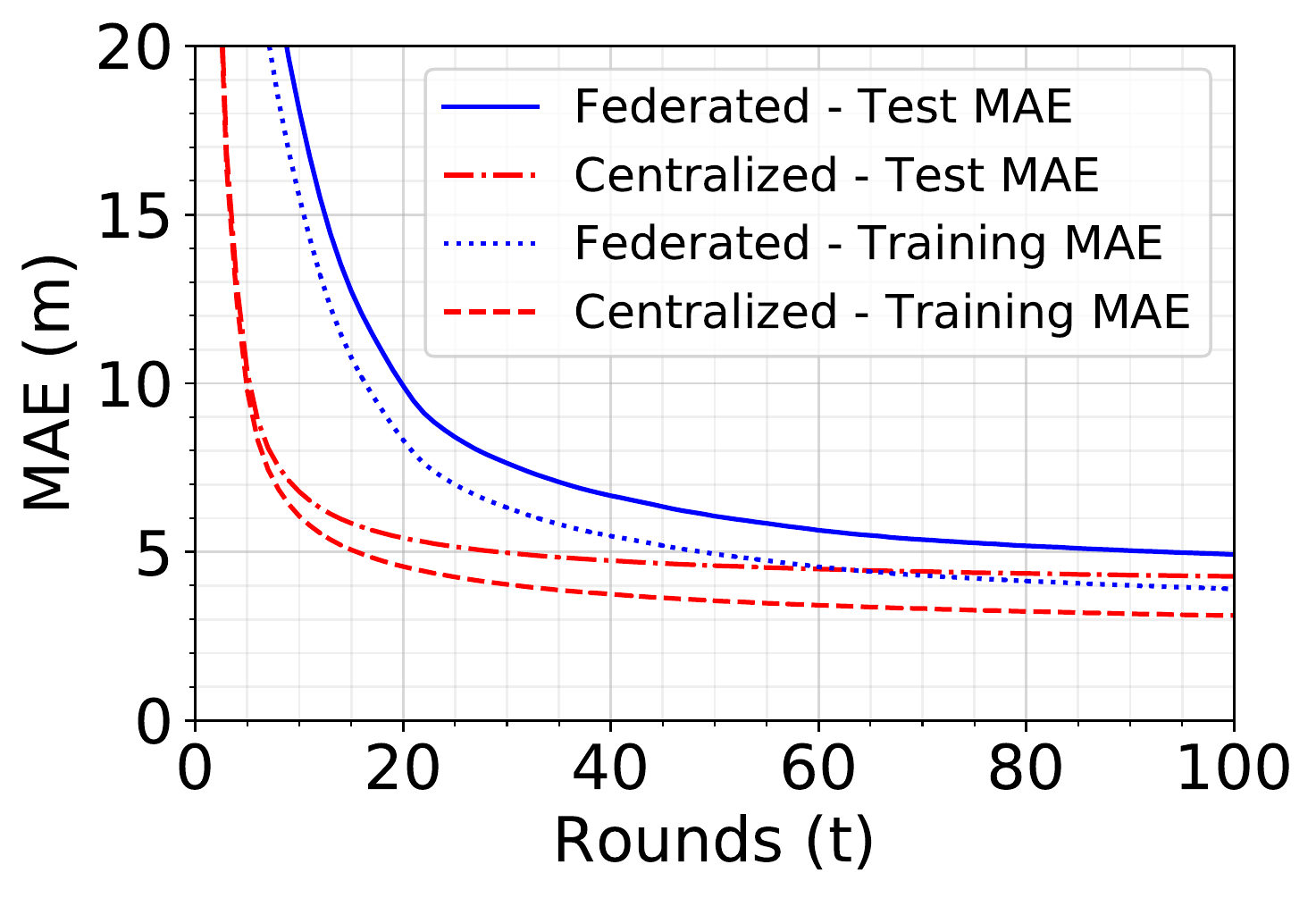}
        \caption{S-3 for n=15}
    \end{subfigure}
    \caption{Effect of number of users and various heterogeneity scenarios on accuracy and convergence time.}
    \label{fig:S123Results}
    \vspace{-5mm}
\end{figure*}

Change in the MAE throughout rounds is shown for Scenario-2 is given in Fig.~\ref{fig:S123Results}(a) to Fig.~\ref{fig:S123Results}(c) with 5, 10, and 15 users.
Uniformly distributed data causes early convergence and overfitting afterward, hence resulting in high training accuracy and lower test accuracy.
For example, the federated model converges around 40 rounds with 15 users, and the gap between training and test accuracy increases afterward for both centralized and federated models.

The effect of location heterogeneity on MAE is presented in Fig.\ref{fig:S123Results}(d) to Fig.~\ref{fig:S123Results}(f).
Note that, the heterogeneity affects both centralized and federated training, as well as increasing the convergence time.
The ramifications of overfitting are less visible compared to the centralized method since federated averaging handles heterogeneous models, which eventually increases variance in the global model, and reduces the bias.
It can be seen that although training accuracy is lower than the uniform case, the test accuracy is higher due to the avoidance of overfitting.
The MAE is $5.53$ with $15$ users.
Heterogeneity in the number of training samples is another challenge for Federated Learning~\cite{mcmahan2016communicationefficient}.
It is shown that increasing heterogeneity increases convergence time and lowers accuracy. However, with enough users, the error eventually drops.

\section{Conclusion}
\label{sect:Conclusion}
In this paper, we propose a federated learning-based technique to provide privacy-preserving crowdsourcing for localization.
The proposed method assures the utilization of the private local data by keeping the data where it is generated by only sharing the local models.
We demonstrate the prominence of federated learning for improving localization accuracy.
We show that the localization accuracy can be improved up to $1.8$ meters using federated learning on top of the centralized dataset.
Federated learning could achieve below $5$ meters MAE localization performance.
We explored the effect of unbalanced sample-size and non-IID spatial heterogeneity of user data on the performance of federated learning.

In our experimental results, we can localize users within $4.98$ meters accuracy in a $390$ meters to $270$ meters building utilizing RSS fingerprint federated learning-based localization.
We are also able to show the performance of the federated learning close to the centralized training (i.e., within $1$ meters for most cases).
In general, we show that federated learning performs well for crowdsourcing of RSS fingerprint for localization in real-world settings while preserving the privacy of the users.
\balance
\bibliographystyle{IEEEtran}
\bibliography{refs}
\end{document}